\documentclass[prb, aps, reprint]{revtex4-1} 
\usepackage{graphicx}
\usepackage{amsmath}
\usepackage{commath}
\usepackage{braket}
\usepackage{hyperref}
\usepackage{siunitx}

\DeclareMathOperator{\tr}{tr}

\begin{document}
\title{Phase pinning and interlayer effects on competing orders in cuprates}
\date{\today}
\author{Zachary M Raines}
\affiliation{Joint Quantum Institute and Condensed Matter Theory Center, Department of Physics, University of Maryland, College Park, Maryland 20742-4111, USA} 
\begin{abstract}
Over the past few years, several exciting experiments in the cuprates have seen evidence of a transient superconducting state upon optical excitation polarized along the c-axis [R. Mankowsky et al., Nature \textbf{516}, 71 (2014)].
The competition between d-form-factor order and superconductivity in these materials has been proposed as an important factor in the observed enhancement of superconductivity.
Central to this effect is the structure of the bond-density-wave along the c-axis, in particular, the $c$-axis component of the ordering vector $Q_z$.
Motivated by the fact that the bond-density-wave order empirically shows a broad peak in c-axis momentum, 
we consider a model of randomly oriented charge ordering domains and study how interlayer coupling affects the competition of this order with superconductivity.
\end{abstract}
\maketitle
\section{Introduction}
The cuprate superconductors have been a topic of active research since their discovery more than thirty years ago\cite{Bednorz1986}.
The past several years have brought exciting new experimental works in underdoped cuprates
on transient states showing signatures of electron-electron pairing\cite{Fausti2011,Kaiser2014,Hu2014,Nicoletti2014}.
In these experiments, the system is excited via mid-infrared laser pulses which
drive phonon modes of the system and can lead to quasi-static changes of the
lattice structure via non-linear phonon couplings\cite{Mankowsky2014a}.
For times close to the pump, features reminiscent of superconductivity can be
seen in the optical conductivity $\sigma(\omega)$, e.g.\ a $1/\omega$ divergence
in $\mathrm{Im}\sigma(\omega)$ and Josephson plasma resonances\cite{Forst2014}.

Also in the last few years, there has been growing interest in charge ordering
phases in several cuprate families\cite{Ghiringhelli2012, Chang2012,
Achkar2012, Coslovich2013,  Fujita2014, Comin2014, Forst2014}, which have been
seen to compete and coexist with superconductivity at low temperatures.
One model for such order is a density-wave instability emerging from nesting of
the Fermi surface\cite{Sau2013,Sachdev2013}.
Such a model predicts the experimentally observed $d$-wave form factor seen in
experiments\cite{Fujita2014}, although it predicts a diagonal $(Q, Q)$
in-plane ordering vector instead of the observed axial $(Q, 0)$
order\cite{Wang2014a}.

The nature of the photo-excitation employed in experiments, as well as previous theoretical works, have suggested that it is important to understand the effect of interlayer coupling. 
Particular its role in the competition between charge order and superconductivity to have a full understanding of the effects seen under mid-infrared excitation. In particular, one scenario suggests melting of the competing charge order\cite{Patel2016a,Raines2015a} via modulation of the interlayer coupling as the underlying mechanism, motivated in particular by the suppression of charge ordering peaks in X-ray coinciding with the transient pairing state\cite{Forst2014}.
Additionally, coupling between the planes seems to play an important role in the experimental results\cite{Hoppner2014a,Mankowsky2014a}.
One consequence of driving the $c$-axis phonon modes is a transient quasistatic modification of the interlayer spacing\cite{Mankowsky2014a},
leading to an enhancement of the hopping between the planes.
In a previous work, we showed that an increase in interlayer coupling could lead to a melting of d-form-factor density wave order in a model of stacked planes\cite{Raines2015a}.
Furthermore, within the model, the melting of the density wave order led to a corresponding enhancement of superconductivity.
These results apply to the case of order which is constant along the $c$-axis, while charge order with $c$-axis momentum $Q_z=\pi$ is robust to changes in the interlayer coupling.
We, however, note that such a configuration is to be contrasted with the empirical observation that while the $c$-axis momentum seen in experiments is peaked about $Q_z=\pi$, the feature is quite broad\cite{Chang2012}.
Along with scanning tunneling microscopy results\cite{Fujita2014}, this suggests a picture of patches of in-plane order which are only weakly correlated between planes. 
We consider here a model where the local phase and orientation of charge order are pinned by e.g.\ lattice impurities or distortions.
Taking this phase and orientation to be random variables, we consider the Landau theory obtained by averaging over all such regions in the system
In general, we find that when in-plane pinning of the charge order is taken into account an increase of interlayer coupling leads to a melting of charge order and an enhancement of superconductivity.

The outline of the paper is as follows.
In Section~\ref{sec:model}, we describe the $t-J-V$ model\cite{Kivelson1990,Dagotto1992} of the planes and
consider the non-interacting susceptibility in the d-form-factor density wave (dFF-DW) channel to find the
wavevector of the strongest instability.
In Section~\ref{sec:review}, we review the mechanism for enhancement of superconductivity in the case where dFF-DW is constant along the $c$-axis.
Then, in Section~\ref{sec:landau}, we consider the averaged Landau free energy of
competing superconductivity and dFF-DW order and study how interlayer coupling affects the competition between the two orders.
Finally, in Section~\ref{sec:conclusion}, we summarize and discuss our results.

\section{Model}
\label{sec:model}
In order to explore the effects of the $c$-axis hopping on d-form-factor density
wave (dFF-DW) order, we consider a minimal model of two planes.
We model each Cu-O plane as a $t-J-V$ model\cite{Kivelson1990,Dagotto1992,Sau2013,Allais2014} on a square lattice, setting the lattice constant $a$ to $1$.
Our model takes the form $H = H_0 + H_\text{int}$.
The free part is given by
\begin{equation}
\begin{gathered}
    H_0 = \int_{\mathbf k}
    \psi^\dagger_{\mathbf k}
    \left[ \left(\xi_{\mathbf k} \hat \Lambda_0 + t_{\mathbf k} \hat \Lambda_1\right) \right] \otimes \hat \sigma_0 \psi_{\mathbf k}\\
    t_{\mathbf k} = t_z {(\cos k_x - \cos k_y)}^2/4
    \label{eq:H0}
    \end{gathered}
\end{equation}
where $\Lambda_i$ are Pauli matrices acting in the layer space and $\sigma_i$ act in the spin space.
$\xi_{\mathbf k}$ includes hopping up to third nearest neighbor\footnote{In this work we used $t_1=\SI {430}{meV}$, $t_2 = -0.32t_1$, $t_3 = -0.5t_2$, and $\mu =-1.1856t_1$.} and $t_k$ describes the hopping between layers\cite{chakravarty_interlayer_1993,Xiang2000}. 
To this we add the layer-local interactions
\begin{equation}
    H_\text{int}
    =
    \frac{1}{2}\sum_{\langle i,j \rangle} \sum_{L}
    \left( V n_{i,L} n_{j,L} + J \mathbf{S}_{i,L} \cdot \mathbf{S}_{j,L} \right).
    \label{eq:Hint}
\end{equation}
where $V$ and $J$ are nearest neighbor Coulomb repulsion and spin exchange, respectively, and $L$ is a layer index.
The $V$ term suppresses $d$-wave superconductivity and functions as a way to tune the relative strength of the two instabilities.

The nearest neighbor form of the interaction allows us to decompose the potential into a sum of factorizable potentials
\begin{gather}
    J_{\vec k - \vec k'} = \frac{1}{2} J \sum_l f^l(k) f^l(k')\\
    V_{\vec k - \vec k'} = \frac{1}{2} V \sum_l f^l(k) f^l(k').
\end{gather}
Here, the functions $f^l(k)$, listed in Tab.~\ref{tab:basis}, form a basis of nearest neighbor in-plane interaction vertices
which transform as representations of $D_4$\cite{Thomson2014}
\begin{table}
    \centering
    \begin{tabular}{@{}lcc}
        \hline 
        $l$&$f^l(\mathbf k)$&Representation\\
        \hline 
        1&$\cos k_x - \cos k_y$&$B_1$\\
        2&$\cos k_x + \cos k_y$&$A_1$\\
        3&$\sin k_x - \sin k_y$&$E$\\
        4&$\sin k_x + \sin k_y$&$E$\\
        \hline 
    \end{tabular}
    \caption{Basis functions for factorization of nearest neighbor interactions categorized by the representation of $D_4$ to which they belong.\label{tab:basis}}
\end{table}
Since we are interested in d-wave superconductivity (dSC) and dFF-DW we will be focusing on the terms containing $f^1(\mathbf{k}) = \cos k_x - \cos k_y$, which correspond to a $d_{x^2-y^2}$-like form factor.
In real space such a form factor corresponds to the case where $x$-links and $y$-links have opposite signs.
Self-energy effects due to interactions in other channels will be assumed to have already been taken into account in the free dispersion.

We may then undertake a decoupling in the dFF-DW and superconducting channels.
Due to the form of the interaction, we consider only layer-local order
parameters.
The superconducting order is taken to be d-wave
and constant along the $c$-axis.

With these restrictions, at the mean-field level, we consider the order parameters
\begin{equation}
    \begin{gathered}
    \phi_L(\mathbf{Q}) = \frac{g_\phi}{2}\sum_{\mathbf{k}, \sigma} f^1(\mathbf{k})\braket{c^\dagger_{\mathbf{k}-\mathbf{Q}/2,\sigma,L} c_{\mathbf{\mathbf{k}+\mathbf{Q}/2},\sigma,L}}\\
    \Delta = \frac{g_\Delta}{4}\sum_{\mathbf{k},L,\sigma,\sigma'} f^1(\mathbf{k})\braket{c_{-\mathbf{k}, L, \sigma} (-i \sigma^2_{\sigma\sigma'})c_{\mathbf{k}, L, \sigma'}}
    \end{gathered}
\end{equation}
where $g_{\phi,\Delta} = \frac{3J}{4} \pm V$ and $\braket{\cdots}$ indicates an ensemble average.


\begin{figure}
    \centering
    \includegraphics[width=\linewidth]{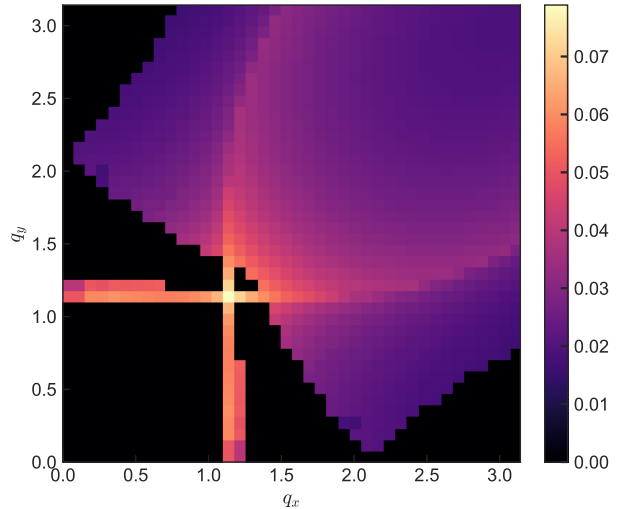}
    \caption{Maximum eigenvalue of the matrix density wave susceptibility $\hat\Pi_\phi$ as a function of in-plane ordering vector in the Brillouin zone.
    The strongest instability is generically at in-plane wavevector $(Q, Q)$ with $Q\sim 1.14$ and out of plane wavevector $Q_z=\pi$.\label{fig:inplanesusc}}
\end{figure}

Having defined the order parameters we can also define associated normal state susceptibilities in these channels.
In particular, we define the matrix dFF-DW susceptibility
\begin{multline}
    \Pi(\mathbf{q})_{ij} = -\sum_k [f^1(\mathbf{k})]^2\\
    \times \tr_{L,\sigma}\left[\hat{G}_0(\epsilon_n, \mathbf{k} + \mathbf{Q}/2)\hat{v}_i\hat{G}_0(\epsilon_n, \mathbf{k} - \mathbf{Q}/2)\hat{v}_j\right].
\end{multline}
where $\hat{G}_0$ is the non-interacting Green's function and $\sum_k$ includes an integral over in-plane momentum and a sum of the Fermionic Matsubara frequency $\epsilon_n$.

In order to determine the in-plane charge-ordering wavevector, we calculated the susceptibility at various values of $\mathbf{Q}$ and compared the maximum eigenvalues.
An intensity plot of the strongest instability by wavevector is shown in Fig.~\ref{fig:inplanesusc}.
For the in-plane component, we generically find the susceptibility to be greatest
for a diagonal $(Q, Q)$ nesting wavevector as is generally the case in such models\cite{Sau2013,Chowdhury2014,Allais2014a}.
\begin{figure}
    \centering
    \includegraphics[width=0.6\linewidth]{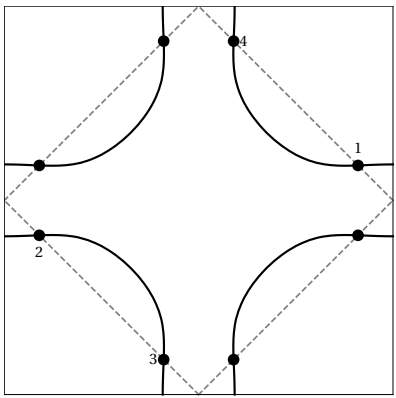}
    \caption{The leading instability has an in-plane ordering momentum which connects the `hotspots', the points where the Fermi surface intersects the magnetic Brillouin zone boundary, across the edge of the Brillouin zone, e.g. the hotspots labeled 1 and 2.
    The symmetry of the problem allows the mean-field Hamiltonian at only hot regions 1 and 2 to be considered.\label{fig:bzhs}}
\end{figure}



\section{Hotspot model and the charge ordering instability}
\label{sec:review}

\begin{figure}
    \centering
    \includegraphics[width=0.7\linewidth]{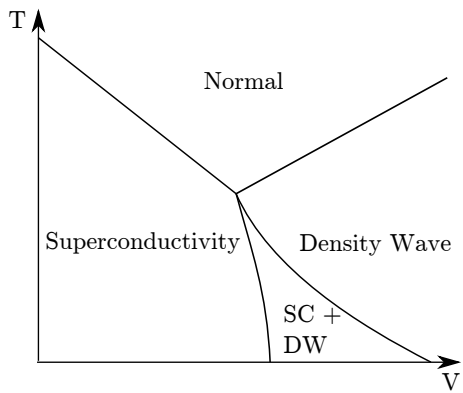}
    \caption{Schematic phase diagram of competing d-wave superconductivity and d-form-factor density wave. $V$, the nearest-neighbor Coulomb repulsion, acts as a tuning parameter for the relative strength of the two instabilities.\label{fig:schematic}}
\end{figure}

As was discussed in a previous work\cite{Raines2015a}, interlayer hopping leads to a curvature effect that suppresses dFF-DW that is constant along the $c$-axis.
This can be understood by looking at an effective low-energy model of `hot-spots' in the Brillouin zone.
We begin by considering a single Cu-O plane.
Noting the importance of anti-ferromagnetic fluctuations in the cuprates, we expand the Hamiltonian about `hot-spots' where the Fermi surface is nested with anti-ferromagnetic wave-vector $(\pi,\pi)$, i.e. where the Fermi surface intersects the magnetic Brillouin zone boundary as depicted in Fig.~\ref{fig:bzhs}.
This leads to a low-energy Hamiltonian
\begin{multline}
        H = \sum_{\vec k,i} \xi_{i,\vec k} c^\dag_{i,\vec k,\sigma}  c_{i,\vec k,\sigma}\\
    + g^{abcd} \sum_{\vec k,\vec p} \left[c^\dag_{1,\vec k,a}c_{2,\vec k,d} c^\dag_{4,\vec p,c} c_{3,\vec p,b}\right.\\
    \left. - c^\dag_{1,\vec k,a} c^\dag_{2,-\vec k,c} c_{4,-\vec p,d} c_{3,\vec p,b}\right],
\end{multline}
where the interaction is
\begin{equation}
    g^{abcd} = -\frac{1}{4}J_{\vec K} \vec{\sigma}_{ab} \cdot \vec{\sigma}_{cd} - V_{\vec K} \delta_{ab} \delta_{cd},
\end{equation}
$\mathbf k$ and $\mathbf p$ are now the deviations from the hotspots, $a-d$ are the electron spin indices, and $i$ is now a hotspot index (e.g. as shown in Fig.~\ref{fig:bzhs}).
Inversion symmetry allows us to restrict attention to half of the hotspots in the Brillouin zone.
We then undertake a simultaneous mean-field decoupling in the dSC and dFF-DW channels.
Due to the symmetries of the problem, it is then only necessary to consider one pair of hotspots with the mean-field Hamiltonian
\begin{equation}
   \hat H_{\text{MF}}(\vec k) =
    \begin{bmatrix}
        \xi_1(\vec k) & \bar \phi & \Delta & 0\\
        \phi & \xi_2(\vec k) & 0 & \Delta\\
        \bar \Delta & 0 & -\xi_1(\vec k) & - \bar \phi\\
         0 &\bar \Delta & - \phi & -\xi_2(\vec k)
    \end{bmatrix},
\end{equation}
where $\Delta$ and $\phi$ are the dSC and dFF-DW order parameters respectively.
From this mean-field Hamiltonian one can obtain a Landau free energy as a function of $g_{\phi,\Delta} = \frac{3J}{4} \pm V$ and temperature:
\begin{equation}
\mathcal{F} = \alpha_\Delta \Delta^2 + \beta_\Delta \Delta^4
+ \alpha_{\phi} \phi^2\\ + \beta_{\phi}\phi^4
+ \gamma \phi^2 \Delta^2.
\label{eq:2dlandau}
\end{equation}
For this model, one finds that $\gamma > 0$, meaning that the orders $\phi$ and $\Delta$ compete.
Nonetheless, there exists a parameter regime where the two coexist, consistent with experimental phase diagrams of the cuprates as schematically depicted in Fig.~\ref{fig:schematic}.

\begin{figure}
    \centering
    \includegraphics[width=0.8\linewidth]{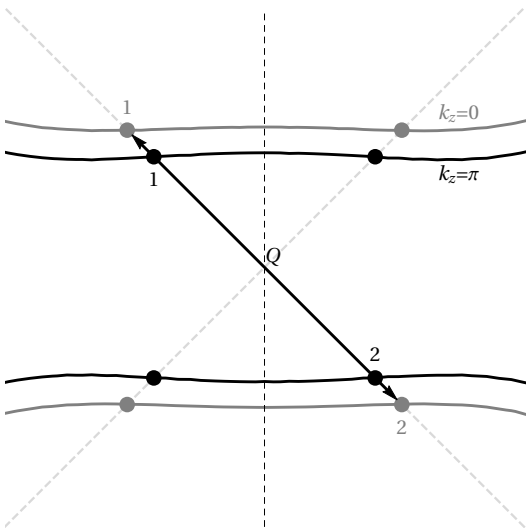}
    \caption{Bending of the Fermi surface as a function of $c$-axis angular momentum leads to a destruction of nesting away from $k_z=0$.
    As the Fermi surface nesting vector is a function of $k_z$ but the ordering vector $\mathbf{Q}$ is not, the Fermi surface cannot remain nested at $\mathbf{Q}$ for all $k_z$.
    This leads to a weakening of the dFF-DW nesting instability.\label{fig:bending}}
\end{figure}

If we now introduce a tunneling between planes along the $c$-axis, the picture is modified in two notable ways.
The most obvious is that the free electron dispersion $\xi$ changes.
However, the dFF-DW order parameter no longer exactly connects hotspots 1 and 2 away from $k_z=0$.
Since the Fermi surface changes shape with $k_z$ while the ordering vector $\mathbf{Q}$ remains fixed, the hotspots cannot be nested at $\mathbf{Q}$ for all values of $k_z$ as can be seen in Fig.~\ref{fig:bending}.
This can most readily be seen by observing the form of the dFF-DW susceptibility within the hotspot model.
If we consider a point, $\mathbf{k}$, the important quantities for the dFF-DW susceptibility are the energies $\xi_\mathbf{k}$ and $\xi_{\mathbf{k} + \mathbf{Q}}$.
Let us define $\xi_\pm = \frac{\xi_\mathbf{k} \pm \xi_{\mathbf{k} + \mathbf{Q}}}{2}.$
We may then express the integrand of the susceptibility as
\begin{equation}
\frac{\sinh{\frac{\xi_-}{2T}}\cosh{\frac{\xi_-}{2T}}}{2\xi_-\left(\sinh^2{\frac{\xi_+}{2T}} + \cosh^2{\frac{\xi_-}{2T}}\right) }.
\end{equation}
From this expression it is clear that $\xi_+$ can only serve to weaken the instability.
In fact, in the case $\xi_+=0$ we recover the logarithmic BCS instability.
In the vicinity of the Fermi surface, where we expect the largest contribution to come from, we can linearize the electron dispersion. 
If the Fermi surface is close to being nested at the dFF-DW wavevector, we have $\xi_+ = \frac{\mathbf{v}_{\mathbf k} \cdot \delta \mathbf{Q}}{2}$ and $\xi_- = \mathbf{v}_{\mathbf{k}} \bar{\mathbf{k}}$, where $\delta \mathbf{Q}$ is the deviation from the perfect nesting wavevector and $\bar{\mathbf{k}}$ is the average of each particle's deviation from the nearest hotspot.
Thus, the extent to which the nesting vector differs from the dFF-DW wavevector determines the strength of the instability.

As tunneling strength is increased, the degree of nesting at the dFF-DW wave-vector is lessened and therefore the instability is weakened.
Effectively, this leads to the value of $\alpha_\phi$ in Eq.~\eqref{eq:2dlandau} increasing.
Superconductivity, on the other hand, is only very weakly affected by the change in coupling strength.
Since the two orders are in competition, the net result is that increasing the interlayer tunneling produces an enhancement of superconductivity within the coexistence regime. For more details we refer the reader to Ref.~\onlinecite{Raines2015a}.

There is, however, an exception to this argument.
Due to the nearest neighbor nature of the inter-plane hopping, the fermi surface is always nested perfectly at the wave vector $(0, 0, \pi)$.
Therefore if the dFF-DW wavevector is $(Q, Q, \pi)$ it will be largely unaffected by the change in interlayer tunneling, and the effect disappears.

The susceptibility analysis of Sec.~\ref{sec:model} indeed confirms, that one can instead have an
instability toward an order which oscillates with wavenumber $\pi$ along the
$c$-axis.
Over a range of parameters, the strongest dFF-DW instability is overwhelmingly of such form.
Nevertheless, there are empirical reasons to believe that the experimental situation is a little more complicated.
The main focus of this work is to address one such aspect.

\section{Effects of phase pinning}
\label{sec:landau}

Empirically, the $c$-axis ordering vector $Q_z$ of the density wave phase is broadly peaked around $\pi$, with a correlation length of approximately $0.6$ lattice units\cite{Chang2012}.
Motivated by this we consider a model in which the interlayer ordering is not defined by a single wavevector.
Instead, we propose a model of the charge order $\phi_{L} = |\phi|e^{i\theta_L}$ at wavevector $\mathbf{Q}_L$, where the relative phase $\theta = \theta_1 - \theta_2$ in between the two layers and the relative orientation $\mathbf{Q}_1 \cdot \mathbf{Q}_2 \in \{0, Q^2\}$ of the ordering vector on the layers are taken to be random variables determined by disorder.

For the model under consideration, the Landau free energy generically takes the form
\begin{multline}
    \mathcal{F}_O[\theta] = \alpha_\Delta |\Delta|^2 + \beta_\Delta |\Delta|^4
    + \alpha_{\phi, O}[\theta] |\phi|^2\\ + \beta_{\phi, O}[\theta] |\phi|^4
    + \gamma_O[\theta] |\phi|^2 |\Delta|^2,
\end{multline}
where $\theta$ is as above, $O = \parallel,\perp$ is the relative orientation of the ordering vectors in the two planes,
and $\Delta$ and $\phi$ are the superconducting and density wave order
parameters, respectively.
The coefficients may be calculated diagrammatically from the free particle action and depend parametrically on the interlayer couplings through the single-particle dispersion.
The microscopic expressions for the Landau coefficients are given in the Appendix. 
We again find $\gamma > 0$, indicating competition between the two orders.
For purposes of calculation it is useful to express the coefficients as a power series in $\cos(\theta)$
\begin{equation}
    \begin{gathered}
    c_\parallel(\theta) = \sum_n c^{(n)} \cos^n \theta\\
    c_\perp = \frac{1}{2\pi}\int_0^{2\pi} \dif \theta c_\parallel(\theta)
    \end{gathered},
    \label{eq:anglevars}
\end{equation}
where $c \in \{\alpha,\beta,\gamma\}$ and for a term including $\phi^m$ the coefficients $c^{(n)} = 0$ for $n > m/2$.
This form allows moments of the terms to be calculated easily in terms of the circular moments $\braket{e^{in\theta}}_\theta$.

The corresponding saddle-point equations admit three non-trivial solutions: a
superconducting phase, a density wave phase, and a coexistent phase:
\begin{equation}
    \label{eq:glsolution}
    \begin{gathered}
    |\Delta| = \sqrt{-\frac{\alpha_\Delta}{2\beta_\Delta}},\ \phi = 0\\
    \Delta = 0,\ |\phi| = \sqrt{-\frac{\alpha_\phi}{2\beta_\phi}}\\
    |\Delta| = \sqrt{\frac{2 \beta_\phi \alpha_\Delta -\gamma \alpha_\phi}{\gamma^2 -4 \beta_\Delta \beta_\phi}},\ |\phi| =
    \sqrt{\frac{2 \beta_\Delta \alpha_\phi -\gamma \alpha_\Delta}{\gamma^2 -4 \beta_\Delta \beta_\phi}}.
    \end{gathered}
\end{equation}

Our goal is to determine how driving the system affects the above phases.
It has been proposed that a significant effect of driving the
apical oxygen modes is a modification of the interlayer spacing.
Following previous works\cite{Nyhus1994,Honma2010,Raines2015a}, we model the effect of this change in spacing on the hoppings as $t_z = t_{z0}\exp(-\alpha d_z)$.
Therefore, we will be interested in seeing how the phase boundaries move as the interlayer hopping strength changes.

\begin{figure}[t]
    \centering
    \includegraphics[width=\linewidth]{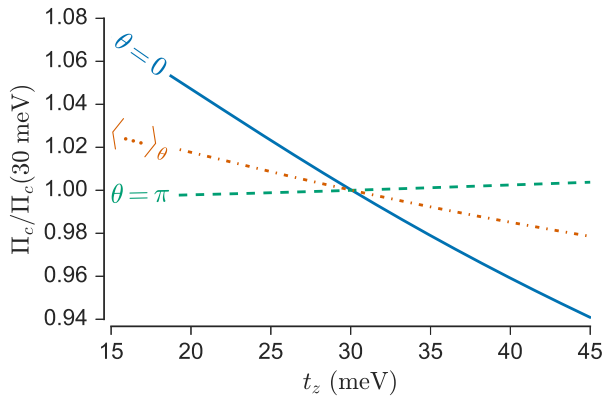}
    \caption{Charge susceptibility as a function of interlayer coupling for relative phase $\theta=0$ (solid blue), $\theta = \pi$ (dashed green), and averaged with respect to $\theta$ (orange dot-dashed).
    Notably, there is little effect for $\theta=0$, while for $\theta=\pi$ there is a noticeable suppression of charge ordering.
    The averaged case sits somewhere between the two, but the suppression of charge order is still significant.
\label{fig:pi_c_theta}}
\end{figure}

To understand the effect of increased $c$-axis coupling on the dFF-DW we first consider the effect on the charge susceptibility $\Pi_\phi = -\alpha_\phi + \frac{1}{g_\phi}$, where $g_\phi$ is the strength of the interaction in the dFF-DW channel.
As can be seen in Fig.~\ref{fig:pi_c_theta}, increasing $t_z$ leads to a notable suppression for order at $\theta = 0$ while $\theta=\pi$ sees a slight enhancement (due to the change in carrier density at fixed chemical potential\cite{Raines2015a}).

Now let us look at the averaged susceptibility.
We take the relative orientation of the wave-vectors to obey a Bernoulli distribution, where alignment has probability $p$, and we take $\theta$ to be distributed according to a wrapped normal distribution\footnote{
The wrapped normal distribution is a straightforward extension of the normal distribution to a periodic variable.
It is a close cousin of the Von Mises distribution, which is the eigendistribution of diffusion for a periodic variable with a harmonic confinement but is somewhat more analytically convenient.
We have explicitly checked that there is no qualitative difference between the results for the two distributions.
} with mean $\mu=\pi$, and standard deviation $\sigma$
\begin{equation}
P[\theta] = \frac{1}{\sigma \sqrt{2\pi}} \sum_{k=-\infty}^{\infty}\exp\left(-\frac{(\theta - \mu + 2\pi k)^2}{2\sigma^2}\right).
\end{equation}
This is the simplest extension of a Gaussian distribution to a periodic variable.
Our choice of distribution corresponds to the approximation that the relative phase is mostly determined by its first and second moments.
Here, we have set the mean of the distribution to $\pi$ to reflect both the fact that this is the energetically favored orientation in absence of disorder and that this is experimentally observed to be the peak ordering vector.

After the averaging process, we find
\begin{multline}
   \overline{\Pi}_\phi
   = \sum_O \int_0^{2\pi} \dif \theta P[\theta] P[O] \Pi_\phi[O, \theta]\\
   =
   \Pi_\phi^{(0)} + \frac{1}{2}\left(\frac{1}{2} + e^{-\sigma^2/2}\cos \mu \right)\Pi_\phi^{(1)}
\end{multline}
with $\Pi^{(i)}$ defined as in Eq.~\eqref{eq:anglevars}.

As is shown in Fig.~\ref{fig:pi_c_theta} the averaged susceptibility, like the $\theta = 0$ case, shows a noticeable decrease as $t_z$ is increased, indicating a melting of charge order.

\begin{figure}[t]
    \centering
    \includegraphics[width=\linewidth]{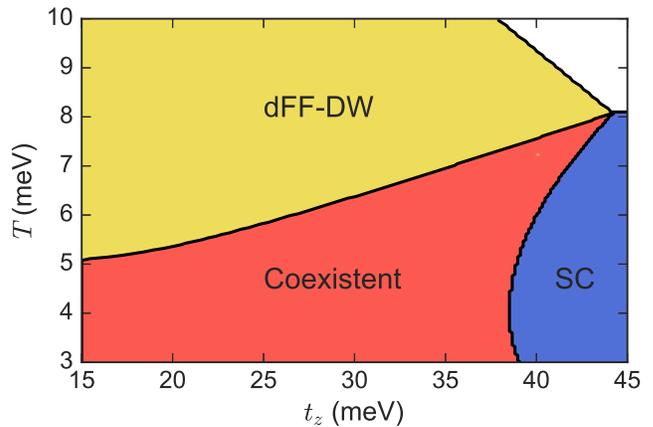}
    \caption{Phase diagram as a function of interlayer coupling $t_z$ and temperature $T$ for a fixed value of interaction strength. Increasing $t_z$ leads to a suppression of charge ordering and a coinciding enhancement of superconductivity.\label{fig:phase}}
\end{figure}

Minimizing the averaged free energy density $\overline{F}$ we find that an increase in interlayer coupling leads to an observable melting of dFF-DW and a concomitant enhancement of superconductivity as can be seen in Fig.~\ref{fig:phase}.
In fact, tuning of interlayer coupling at fixed temperature can tune between charge-ordered, coexistent, and superconducting phases.

\begin{figure}
    \centering
    \includegraphics[width=\linewidth]{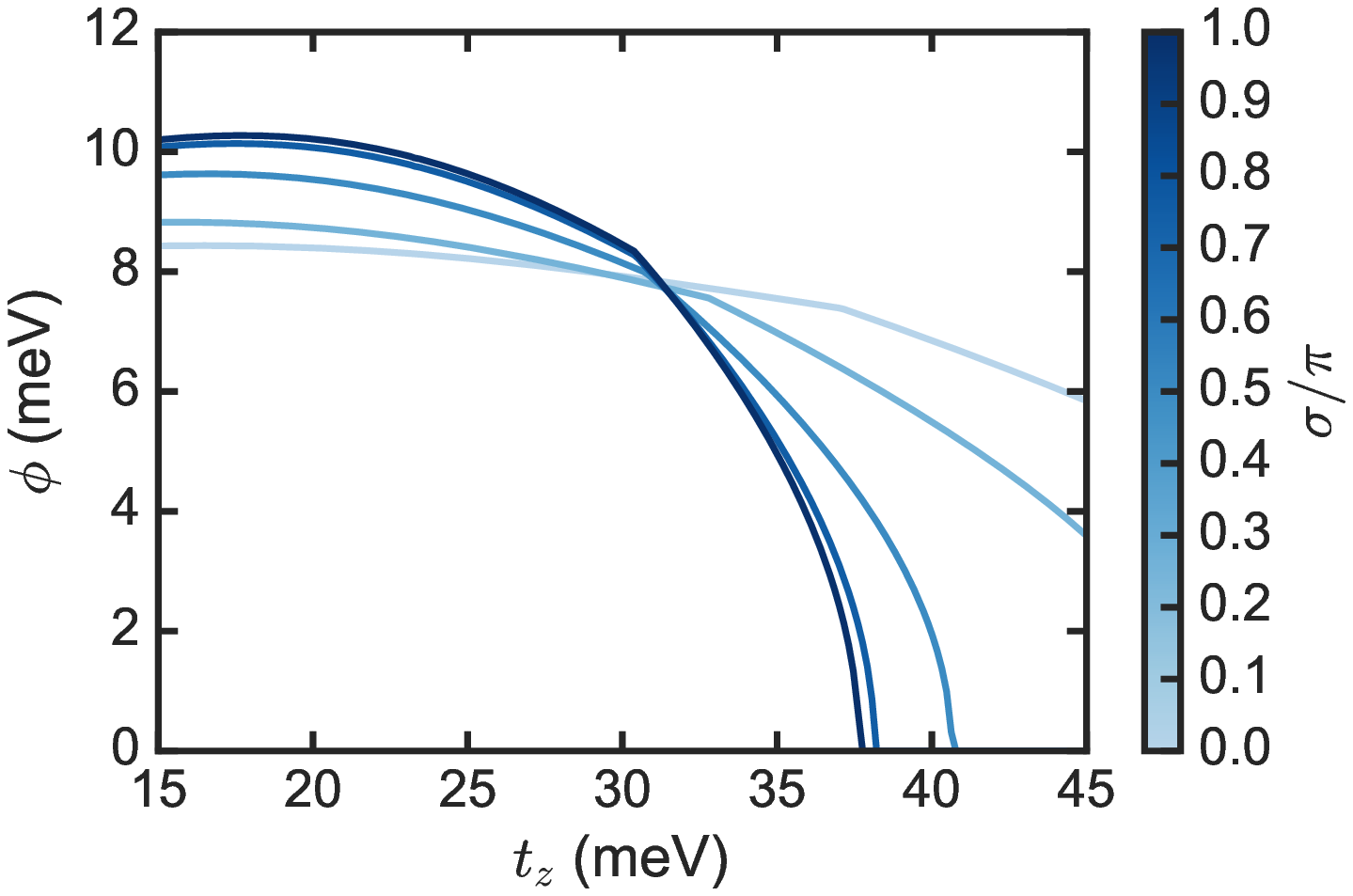}
    \includegraphics[width=\linewidth]{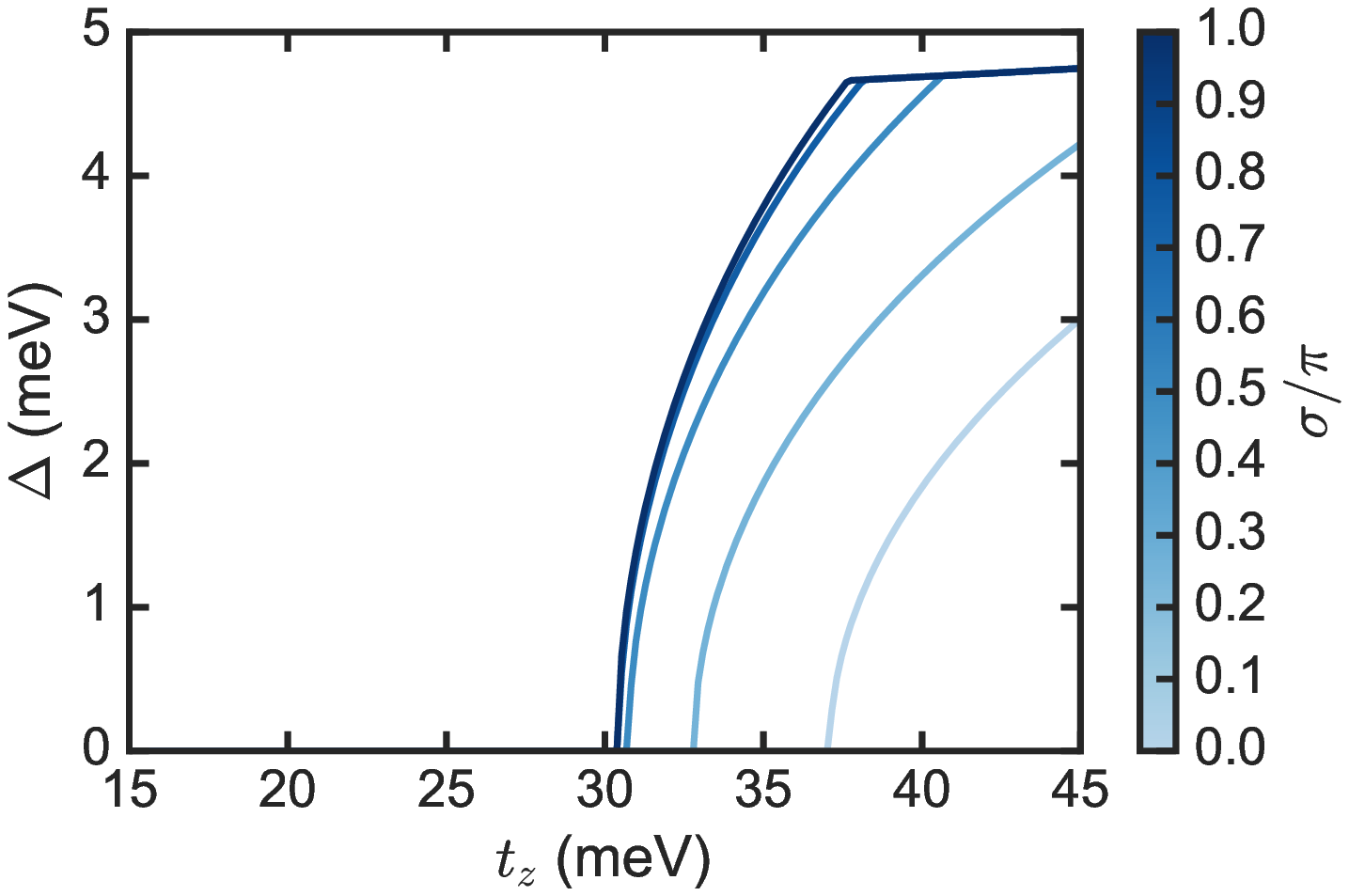}
    \caption{dFF-DW order parameter $\phi$ (top) and superconducting order parameter $\Delta$ (bottom) vs interlayer coupling for various values of $\sigma$, the standard deviation of the interlayer dFF-DW phase difference.
    The coupling constants of the model have been normalized to keep the bare charge ordering temperature at $t_z = \SI{30}{mev}$ fixed.
    Increasing $t_z$ in general leads to a melting of dFF-DW and enhancement of SC with the effect becoming more pronounced as $\sigma$ is increased.
    \label{fig:vssigma}}
\end{figure}

Changing the variance of the phase leads to a quantitative difference but results are qualitatively similar.
In particular, we considered various values of $\sigma$ with the dFF-DW ordering temperature at $t_z = \SI{30}{meV}$ held fixed.
As shown in Fig.~\ref{fig:vssigma} for a wide range of $\sigma$ an increase in interlayer tunneling leads to a melting of dFF-DW and an associated enhancement of dSC.
The salient point is that pinning of the dFF-DW phase in general frustrates the interlayer ordering of the density wave state which would otherwise make it insensitive to changes in interlayer coupling.
So while, in an idealized system the interlayer coupling strength should not appreciably affect the competition between dFF-DW and dSC order, in a realistic system an increase of the interlayer coupling generically leads to a melting of dFF-DW and enhancement of dSC. 

\section{Discussion and Conclusion}
\label{sec:conclusion}
In this work, we have shown that the broadly peaked nature of the $c$-axis structure for dFF-DW order in general means that that an increase of interlayer tunneling leads to a melting of charge order and a corresponding enhancement of the competing superconductivity.
Notably the presence of phase pinning or the dFF-DW is essential to this effect.
This has implications for optical control experiments, which are believed to be inducing a transient superconducting state by coupling to interlayer degrees of freedom, and, in particular, to our previous work in which we investigated the effect of interlayer separation in cuprates on the competition between superconductivity and dFF-DW\cite{Raines2015a}.
In that work we considered specifically the case of an order which is constant along the $c$-axis.
Here we have extended our analysis to consider the case where the interlayer phase difference for the dFF-DW order is a random variable to be averaged over.

One way to visualize our results is that domains with distinct phase differences form and these domains are susceptible to melting to different degrees.
Such a picture is consistent with experiments where inhomogeneous enhancement of electron-electron pairing is observed\cite{Mankowsky2014a}, as one might expect from inhomogeneous melting of dFF-DW domains.

While the $c$-axis curvature effects seem to be too weak to explain the observed enhancement of superconducting correlations alone, there are still other theoretical\cite{Patel2016a} and experimental\cite{Forst2014} reasons to believe that melting of charge order plays an important role.
Other explanations have been considered for this effect such as redistribution of spectral weight\cite{Hoppner2014a}, suppression of superconducting phase fluctuations\cite{denny_proposed_2015}, or other routes to melting of dFF-DW order\cite{Patel2016a}.
Most likely the complete explanation is some combination of factors, with a number of these frameworks forming complementary rather than competing mechanisms.

\begin{acknowledgments}
We would like to thank Victor Galitski and Valentin Stanev for helpful discussions.
This work was supported by NSF DMR-1613029 and the Simons Foundation.
\end{acknowledgments}

\appendix*
\section{Microscopic expressions for Landau coefficients}
\label{sec:landau-coeff}

As discussed in Sec.~\ref{sec:landau}, the Landau theory for competing orders in this model takes the form
\begin{multline}
    \mathcal{F}_O[\theta] = \alpha_\Delta |\Delta|^2 + \beta_\Delta |\Delta|^4
    + \alpha_{\phi, O}[\theta] |\phi|^2\\ + \beta_{\phi, O}[\theta] |\phi|^4
    + \gamma_O[\theta] |\phi|^2 |\Delta|^2,
\end{multline}

The quadratic Landau coefficients are simply related to the susceptibilities
in the corresponding channels
\begin{equation}
    \begin{gathered}
        \alpha_\Delta = \frac{1}{g_\Delta} - \Pi_\Delta\\
        \alpha_\phi = \frac{1}{g_\phi} - \Pi_\phi
    \end{gathered}.
\end{equation}
The superconducting terms are simplest, with
\begin{equation}
    \Pi_\Delta = \sum_{\mathbf k} f^1{(\mathbf k)}^2
    \sum_\pm
    \frac{\tanh{\frac{\epsilon_{\mathbf{k}, \pm}}{2T}}}
    {2 \epsilon_{\mathbf{k}, \pm}}
\end{equation}
and
\begin{multline}
    \beta_\Delta = \sum_{\mathbf k} f^1{(\mathbf k)}^2
    \sum_\pm \frac{1}{2\epsilon_{\mathbf{k}, \pm}}\\
    \times
    \left[
    \frac{\tanh{\frac{\epsilon_{\mathbf{k}, \pm}}{2T}}}
    {2 \epsilon_{\mathbf{k}, \pm}}
    + n'_f(\epsilon_{\mathbf{k}, \pm})
    \right],
\end{multline}
where $\epsilon_{\mathbf{k}, \pm} = \xi_{\mathbf k} \pm t_{\mathbf k}$ are the eigenvalues of the free Hamiltonian.

As mentioned above the terms involving $\phi$ can be broken into coefficients of $\cos^n \theta$.
Beginning with the dFF-DW susceptibility
\begin{equation}
    \Pi^{(n)}_\phi = \sum_{\mathbf k} f^1{(\mathbf k)}^2
    \sum_{\lambda\lambda'} {(-1)}^{n(\lambda - \lambda')}\Pi(\epsilon_{\mathbf{k}, \lambda}, \epsilon_{\mathbf{k}, \lambda'})
\end{equation}
where
\begin{equation}
    \Pi(\epsilon_1, \epsilon_2) = \frac{n_f(\epsilon_2) - n_f(\epsilon_1)}{\epsilon_1 - \epsilon_2}.
\end{equation}

\begin{widetext}
The quartic dFF-DW terms are
\begin{multline}
    \beta_\phi^{(1)} = \sum_{\mathbf k}\sum_{\lambda\lambda'}
    {(-1)}^{\lambda -\lambda'}
    \left[
    \frac{1}{2}f^1{(\mathbf k)}^4M_{\beta,1}(\epsilon_\lambda(\mathbf{k} + \mathbf{Q}{2}), \epsilon_{\lambda'}(\mathbf{k} - \mathbf{Q}{2}))\right.\\
        \left.
        +f^1{(\mathbf{k} + \mathbf{Q})}^2f^1{(\mathbf{k} - \mathbf{Q})}^2
        M_{\beta,2}(\epsilon_\lambda(\mathbf k),
            \epsilon_{\lambda'}(\mathbf k - \mathbf{Q}),
        \epsilon_{\lambda'}(\mathbf k + \mathbf{Q}))
    \right]
\end{multline} and
\begin{multline}
    \beta_\phi^{(0,2)} =
    \sum_{\mathbf k}
    \left\{
        f^1{(\mathbf k)}^4 \left[
            \frac{1}{4}\sum_{\lambda\lambda'} M_{\beta,1}(\epsilon_\lambda(\mathbf{k} + \mathbf{Q}{2}), \epsilon_{\lambda'}(\mathbf{k} - \mathbf{Q}{2}))\right.\right.\\
            \left.
            \mp
            M_{\beta,3}(\{ \epsilon_{\lambda}(\mathbf k +\zeta \mathbf Q)\}_{\lambda,\zeta})
            \pm \sum_\lambda M_{\beta,2}(\epsilon_\lambda(\mathbf k - \mathbf{Q}/2),
                \epsilon_{\lambda'}(\mathbf k + \mathbf{Q}/2),
            \epsilon_{-\lambda}(\mathbf k + \mathbf{Q}/2))
        \right]\\
        + f^1{(\mathbf{k} + \mathbf{Q})}^2 f^1{(\mathbf{k} - \mathbf{Q})}^2
        \left[
            \frac{1}{2}\sum_{\lambda_1,\lambda_2,\lambda_3}{(\pm1)}^{\lambda_2 - \lambda_3}
            M_{\beta,2}(\epsilon_{\lambda_1}(\mathbf k),
                \epsilon_{\lambda_2}(\mathbf k - \mathbf{Q}),
            \epsilon_{\lambda_3}(\mathbf k + \mathbf{Q}))\right.\\
            \left.\left.
            \mp \sum_{\lambda\lambda'}{(-1)}^{\lambda - \lambda'}
            M_{\beta,3}(\{\epsilon_{+}(\mathbf k),
                    \epsilon_{-}(\mathbf k),
                    \epsilon_{\lambda}(\mathbf k + \mathbf Q),
            \epsilon_{\lambda'}(\mathbf k - \mathbf Q)\})
        \right]
    \right\},
\end{multline}
where we have defined
\begin{equation}
\begin{gathered}
    M_{\beta, 1}(x, y) =
    \frac{1}{{(x-y)}^2}
    \left(
        \frac{\tanh{\frac{x}{2T}} - \tanh{\frac{y}{2T}}}{x-y}
        + n_f'(x) + n_f'(y)
    \right)
    \\
    M_{\beta, 2}(x, y, z) = \frac{1}{z-y} \left(
        \frac{n_f(z)}{{(x-z)}^2} - \frac{n_f(y)}{{(x-y)}^2}
    \right)
    +
    \frac{1}{(x-z)(x-y)}
    \left[
        n_f'(x) - n_f(x)
        \left(\frac{1}{x-z} + \frac{1}{x-y}\right)
    \right]
    \\
    M_{\beta, 3}(\{x_i\}) = \sum_i \prod_{j\neq i} \frac{n_f(x_i)}{x_i - x_j}.
\end{gathered}
\end{equation}
\end{widetext}

Finally for the competition term
\begin{multline}
    \gamma^{(n)} =
    \sum_{\mathbf k}
    f^1{(\mathbf k)}^2 f^1(\mathbf{k}_+)
    \sum_{\lambda\lambda'}\\
    \times
    {(-1)}^{n(\lambda - \lambda')}
    \left[
         f^1(\mathbf{k}_+)
         M_{\gamma, 2}(\epsilon_\lambda(\mathbf{k}_+),
     \epsilon_\lambda(\mathbf{k}_-))\right.\\
        \left.
        - f^1(\mathbf{k}_-)
         M_{\gamma, 1}(\epsilon_\lambda(\mathbf{k}_+),
         \epsilon_\lambda(\mathbf{k}_-))
    \right]
\end{multline}
where $\mathbf{k}_\pm = \mathbf{k} \pm \mathbf{Q}/2$ and we have defined
\begin{equation}
\begin{gathered}
    \label{eq:gamma12}
    M_{\gamma, 1}(x, y)  = \frac{1}{{(x)}^2 - {(y)}^2} \left( \frac{\tanh \frac{y}{2T}}{2y} -
    \frac{\tanh \frac{x}{2T}}{2x}\right)\\
    M_{\gamma,2}(x, y) = \frac{1}{2x} \left[
    \frac{n'_f(x)}{x-y} - \frac{\tanh(\frac{x}{2T})}{2x(x+y)} \right]
    + (x \leftrightarrow y).
\end{gathered}
\end{equation}
\bibliography{references}

\begin{thebibliography}{32}%
\makeatletter
\providecommand \@ifxundefined [1]{%
 \@ifx{#1\undefined}
}%
\providecommand \@ifnum [1]{%
 \ifnum #1\expandafter \@firstoftwo
 \else \expandafter \@secondoftwo
 \fi
}%
\providecommand \@ifx [1]{%
 \ifx #1\expandafter \@firstoftwo
 \else \expandafter \@secondoftwo
 \fi
}%
\providecommand \natexlab [1]{#1}%
\providecommand \enquote  [1]{``#1''}%
\providecommand \bibnamefont  [1]{#1}%
\providecommand \bibfnamefont [1]{#1}%
\providecommand \citenamefont [1]{#1}%
\providecommand \href@noop [0]{\@secondoftwo}%
\providecommand \href [0]{\begingroup \@sanitize@url \@href}%
\providecommand \@href[1]{\@@startlink{#1}\@@href}%
\providecommand \@@href[1]{\endgroup#1\@@endlink}%
\providecommand \@sanitize@url [0]{\catcode `\\12\catcode `\$12\catcode
  `\&12\catcode `\#12\catcode `\^12\catcode `\_12\catcode `\%12\relax}%
\providecommand \@@startlink[1]{}%
\providecommand \@@endlink[0]{}%
\providecommand \url  [0]{\begingroup\@sanitize@url \@url }%
\providecommand \@url [1]{\endgroup\@href {#1}{\urlprefix }}%
\providecommand \urlprefix  [0]{URL }%
\providecommand \Eprint [0]{\href }%
\providecommand \doibase [0]{http://dx.doi.org/}%
\providecommand \selectlanguage [0]{\@gobble}%
\providecommand \bibinfo  [0]{\@secondoftwo}%
\providecommand \bibfield  [0]{\@secondoftwo}%
\providecommand \translation [1]{[#1]}%
\providecommand \BibitemOpen [0]{}%
\providecommand \bibitemStop [0]{}%
\providecommand \bibitemNoStop [0]{.\EOS\space}%
\providecommand \EOS [0]{\spacefactor3000\relax}%
\providecommand \BibitemShut  [1]{\csname bibitem#1\endcsname}%
\let\auto@bib@innerbib\@empty
\bibitem [{\citenamefont {Bednorz}\ and\ \citenamefont
  {M{\"{u}}ller}(1986)}]{Bednorz1986}%
  \BibitemOpen
  \bibfield  {author} {\bibinfo {author} {\bibfnamefont {J.~G.}\ \bibnamefont
  {Bednorz}}\ and\ \bibinfo {author} {\bibfnamefont {K.~A.}\ \bibnamefont
  {M{\"{u}}ller}},\ }\href {\doibase 10.1007/BF01303701} {\bibfield  {journal}
  {\bibinfo  {journal} {Z. Phys. B Condens. Matter}\ }\textbf {\bibinfo
  {volume} {64}},\ \bibinfo {pages} {189} (\bibinfo {year} {1986})}\BibitemShut
  {NoStop}%
\bibitem [{\citenamefont {Fausti}\ \emph {et~al.}(2011)\citenamefont {Fausti},
  \citenamefont {Tobey}, \citenamefont {Dean}, \citenamefont {Kaiser},
  \citenamefont {Dienst}, \citenamefont {Hoffmann}, \citenamefont {Pyon},
  \citenamefont {Takayama}, \citenamefont {Takagi},\ and\ \citenamefont
  {Cavalleri}}]{Fausti2011}%
  \BibitemOpen
  \bibfield  {author} {\bibinfo {author} {\bibfnamefont {D.}~\bibnamefont
  {Fausti}}, \bibinfo {author} {\bibfnamefont {R.~I.}\ \bibnamefont {Tobey}},
  \bibinfo {author} {\bibfnamefont {N.}~\bibnamefont {Dean}}, \bibinfo {author}
  {\bibfnamefont {S.}~\bibnamefont {Kaiser}}, \bibinfo {author} {\bibfnamefont
  {A.}~\bibnamefont {Dienst}}, \bibinfo {author} {\bibfnamefont {M.~C.}\
  \bibnamefont {Hoffmann}}, \bibinfo {author} {\bibfnamefont {S.}~\bibnamefont
  {Pyon}}, \bibinfo {author} {\bibfnamefont {T.}~\bibnamefont {Takayama}},
  \bibinfo {author} {\bibfnamefont {H.}~\bibnamefont {Takagi}}, \ and\ \bibinfo
  {author} {\bibfnamefont {A.}~\bibnamefont {Cavalleri}},\ }\href {\doibase
  10.1126/science.1197294} {\bibfield  {journal} {\bibinfo  {journal}
  {Science}\ }\textbf {\bibinfo {volume} {331}},\ \bibinfo {pages} {189}
  (\bibinfo {year} {2011})}\BibitemShut {NoStop}%
\bibitem [{\citenamefont {Kaiser}\ \emph {et~al.}(2014)\citenamefont {Kaiser},
  \citenamefont {Hunt}, \citenamefont {Nicoletti}, \citenamefont {Hu},
  \citenamefont {Gierz}, \citenamefont {Liu}, \citenamefont {{Le Tacon}},
  \citenamefont {Loew}, \citenamefont {Haug}, \citenamefont {Keimer},\ and\
  \citenamefont {Cavalleri}}]{Kaiser2014}%
  \BibitemOpen
  \bibfield  {author} {\bibinfo {author} {\bibfnamefont {S.}~\bibnamefont
  {Kaiser}}, \bibinfo {author} {\bibfnamefont {C.~R.}\ \bibnamefont {Hunt}},
  \bibinfo {author} {\bibfnamefont {D.}~\bibnamefont {Nicoletti}}, \bibinfo
  {author} {\bibfnamefont {W.}~\bibnamefont {Hu}}, \bibinfo {author}
  {\bibfnamefont {I.}~\bibnamefont {Gierz}}, \bibinfo {author} {\bibfnamefont
  {H.~Y.}\ \bibnamefont {Liu}}, \bibinfo {author} {\bibfnamefont
  {M.}~\bibnamefont {{Le Tacon}}}, \bibinfo {author} {\bibfnamefont
  {T.}~\bibnamefont {Loew}}, \bibinfo {author} {\bibfnamefont {D.}~\bibnamefont
  {Haug}}, \bibinfo {author} {\bibfnamefont {B.}~\bibnamefont {Keimer}}, \ and\
  \bibinfo {author} {\bibfnamefont {A.}~\bibnamefont {Cavalleri}},\ }\href
  {\doibase 10.1103/PhysRevB.89.184516} {\bibfield  {journal} {\bibinfo
  {journal} {Phys. Rev. B}\ }\textbf {\bibinfo {volume} {89}},\ \bibinfo
  {pages} {184516} (\bibinfo {year} {2014})}\BibitemShut {NoStop}%
\bibitem [{\citenamefont {Hu}\ \emph {et~al.}(2014)\citenamefont {Hu},
  \citenamefont {Kaiser}, \citenamefont {Nicoletti}, \citenamefont {Hunt},
  \citenamefont {Gierz}, \citenamefont {Hoffmann}, \citenamefont {{Le Tacon}},
  \citenamefont {Loew}, \citenamefont {Keimer},\ and\ \citenamefont
  {Cavalleri}}]{Hu2014}%
  \BibitemOpen
  \bibfield  {author} {\bibinfo {author} {\bibfnamefont {W.}~\bibnamefont
  {Hu}}, \bibinfo {author} {\bibfnamefont {S.}~\bibnamefont {Kaiser}}, \bibinfo
  {author} {\bibfnamefont {D.}~\bibnamefont {Nicoletti}}, \bibinfo {author}
  {\bibfnamefont {C.~R.}\ \bibnamefont {Hunt}}, \bibinfo {author}
  {\bibfnamefont {I.}~\bibnamefont {Gierz}}, \bibinfo {author} {\bibfnamefont
  {M.~C.}\ \bibnamefont {Hoffmann}}, \bibinfo {author} {\bibfnamefont
  {M.}~\bibnamefont {{Le Tacon}}}, \bibinfo {author} {\bibfnamefont
  {T.}~\bibnamefont {Loew}}, \bibinfo {author} {\bibfnamefont {B.}~\bibnamefont
  {Keimer}}, \ and\ \bibinfo {author} {\bibfnamefont {A.}~\bibnamefont
  {Cavalleri}},\ }\href {\doibase 10.1038/nmat3963} {\bibfield  {journal}
  {\bibinfo  {journal} {Nat. Mater.}\ }\textbf {\bibinfo {volume} {13}},\
  \bibinfo {pages} {705} (\bibinfo {year} {2014})}\BibitemShut {NoStop}%
\bibitem [{\citenamefont {Nicoletti}\ \emph {et~al.}(2014)\citenamefont
  {Nicoletti}, \citenamefont {Casandruc}, \citenamefont {Laplace},
  \citenamefont {Khanna}, \citenamefont {Hunt}, \citenamefont {Kaiser},
  \citenamefont {Dhesi}, \citenamefont {Gu}, \citenamefont {Hill},\ and\
  \citenamefont {Cavalleri}}]{Nicoletti2014}%
  \BibitemOpen
  \bibfield  {author} {\bibinfo {author} {\bibfnamefont {D.}~\bibnamefont
  {Nicoletti}}, \bibinfo {author} {\bibfnamefont {E.}~\bibnamefont
  {Casandruc}}, \bibinfo {author} {\bibfnamefont {Y.}~\bibnamefont {Laplace}},
  \bibinfo {author} {\bibfnamefont {V.}~\bibnamefont {Khanna}}, \bibinfo
  {author} {\bibfnamefont {C.~R.}\ \bibnamefont {Hunt}}, \bibinfo {author}
  {\bibfnamefont {S.}~\bibnamefont {Kaiser}}, \bibinfo {author} {\bibfnamefont
  {S.~S.}\ \bibnamefont {Dhesi}}, \bibinfo {author} {\bibfnamefont {G.~D.}\
  \bibnamefont {Gu}}, \bibinfo {author} {\bibfnamefont {J.~P.}\ \bibnamefont
  {Hill}}, \ and\ \bibinfo {author} {\bibfnamefont {A.}~\bibnamefont
  {Cavalleri}},\ }\href {\doibase 10.1103/PhysRevB.90.100503} {\bibfield
  {journal} {\bibinfo  {journal} {Phys. Rev. B}\ }\textbf {\bibinfo {volume}
  {90}},\ \bibinfo {pages} {100503} (\bibinfo {year} {2014})}\BibitemShut
  {NoStop}%
\bibitem [{\citenamefont {Mankowsky}\ \emph {et~al.}(2014)\citenamefont
  {Mankowsky}, \citenamefont {Subedi}, \citenamefont {F{\"{o}}rst},
  \citenamefont {Mariager}, \citenamefont {Chollet}, \citenamefont {Lemke},
  \citenamefont {Robinson}, \citenamefont {Glownia}, \citenamefont {Minitti},
  \citenamefont {Frano}, \citenamefont {Fechner}, \citenamefont {Spaldin},
  \citenamefont {Loew}, \citenamefont {Keimer}, \citenamefont {Georges},\ and\
  \citenamefont {Cavalleri}}]{Mankowsky2014a}%
  \BibitemOpen
  \bibfield  {author} {\bibinfo {author} {\bibfnamefont {R.}~\bibnamefont
  {Mankowsky}}, \bibinfo {author} {\bibfnamefont {A.}~\bibnamefont {Subedi}},
  \bibinfo {author} {\bibfnamefont {M.}~\bibnamefont {F{\"{o}}rst}}, \bibinfo
  {author} {\bibfnamefont {S.~O.}\ \bibnamefont {Mariager}}, \bibinfo {author}
  {\bibfnamefont {M.}~\bibnamefont {Chollet}}, \bibinfo {author} {\bibfnamefont
  {H.~T.}\ \bibnamefont {Lemke}}, \bibinfo {author} {\bibfnamefont {J.~S.}\
  \bibnamefont {Robinson}}, \bibinfo {author} {\bibfnamefont {J.~M.}\
  \bibnamefont {Glownia}}, \bibinfo {author} {\bibfnamefont {M.~P.}\
  \bibnamefont {Minitti}}, \bibinfo {author} {\bibfnamefont {A.}~\bibnamefont
  {Frano}}, \bibinfo {author} {\bibfnamefont {M.}~\bibnamefont {Fechner}},
  \bibinfo {author} {\bibfnamefont {N.~A.}\ \bibnamefont {Spaldin}}, \bibinfo
  {author} {\bibfnamefont {T.}~\bibnamefont {Loew}}, \bibinfo {author}
  {\bibfnamefont {B.}~\bibnamefont {Keimer}}, \bibinfo {author} {\bibfnamefont
  {A.}~\bibnamefont {Georges}}, \ and\ \bibinfo {author} {\bibfnamefont
  {A.}~\bibnamefont {Cavalleri}},\ }\href {\doibase 10.1038/nature13875}
  {\bibfield  {journal} {\bibinfo  {journal} {Nature}\ }\textbf {\bibinfo
  {volume} {516}},\ \bibinfo {pages} {71} (\bibinfo {year} {2014})}\BibitemShut
  {NoStop}%
\bibitem [{\citenamefont {F{\"{o}}rst}\ \emph {et~al.}(2014)\citenamefont
  {F{\"{o}}rst}, \citenamefont {Frano}, \citenamefont {Kaiser}, \citenamefont
  {Mankowsky}, \citenamefont {Hunt}, \citenamefont {Turner}, \citenamefont
  {Dakovski}, \citenamefont {Minitti}, \citenamefont {Robinson}, \citenamefont
  {Loew}, \citenamefont {{Le Tacon}}, \citenamefont {Keimer}, \citenamefont
  {Hill}, \citenamefont {Cavalleri},\ and\ \citenamefont {Dhesi}}]{Forst2014}%
  \BibitemOpen
  \bibfield  {author} {\bibinfo {author} {\bibfnamefont {M.}~\bibnamefont
  {F{\"{o}}rst}}, \bibinfo {author} {\bibfnamefont {A.}~\bibnamefont {Frano}},
  \bibinfo {author} {\bibfnamefont {S.}~\bibnamefont {Kaiser}}, \bibinfo
  {author} {\bibfnamefont {R.}~\bibnamefont {Mankowsky}}, \bibinfo {author}
  {\bibfnamefont {C.~R.}\ \bibnamefont {Hunt}}, \bibinfo {author}
  {\bibfnamefont {J.~J.}\ \bibnamefont {Turner}}, \bibinfo {author}
  {\bibfnamefont {G.~L.}\ \bibnamefont {Dakovski}}, \bibinfo {author}
  {\bibfnamefont {M.~P.}\ \bibnamefont {Minitti}}, \bibinfo {author}
  {\bibfnamefont {J.}~\bibnamefont {Robinson}}, \bibinfo {author}
  {\bibfnamefont {T.}~\bibnamefont {Loew}}, \bibinfo {author} {\bibfnamefont
  {M.}~\bibnamefont {{Le Tacon}}}, \bibinfo {author} {\bibfnamefont
  {B.}~\bibnamefont {Keimer}}, \bibinfo {author} {\bibfnamefont {J.~P.}\
  \bibnamefont {Hill}}, \bibinfo {author} {\bibfnamefont {A.}~\bibnamefont
  {Cavalleri}}, \ and\ \bibinfo {author} {\bibfnamefont {S.~S.}\ \bibnamefont
  {Dhesi}},\ }\href {\doibase 10.1103/PhysRevB.90.184514} {\bibfield  {journal}
  {\bibinfo  {journal} {Phys. Rev. B}\ }\textbf {\bibinfo {volume} {90}},\
  \bibinfo {pages} {184514} (\bibinfo {year} {2014})}\BibitemShut {NoStop}%
\bibitem [{\citenamefont {Ghiringhelli}\ \emph {et~al.}(2012)\citenamefont
  {Ghiringhelli}, \citenamefont {{Le Tacon}}, \citenamefont {Minola},
  \citenamefont {Blanco-Canosa}, \citenamefont {Mazzoli}, \citenamefont
  {Brookes}, \citenamefont {{De Luca}}, \citenamefont {Frano}, \citenamefont
  {Hawthorn}, \citenamefont {He}, \citenamefont {Loew}, \citenamefont {Sala},
  \citenamefont {Peets}, \citenamefont {Salluzzo}, \citenamefont {Schierle},
  \citenamefont {Sutarto}, \citenamefont {Sawatzky}, \citenamefont {Weschke},
  \citenamefont {Keimer},\ and\ \citenamefont {Braicovich}}]{Ghiringhelli2012}%
  \BibitemOpen
  \bibfield  {author} {\bibinfo {author} {\bibfnamefont {G.}~\bibnamefont
  {Ghiringhelli}}, \bibinfo {author} {\bibfnamefont {M.}~\bibnamefont {{Le
  Tacon}}}, \bibinfo {author} {\bibfnamefont {M.}~\bibnamefont {Minola}},
  \bibinfo {author} {\bibfnamefont {S.}~\bibnamefont {Blanco-Canosa}}, \bibinfo
  {author} {\bibfnamefont {C.}~\bibnamefont {Mazzoli}}, \bibinfo {author}
  {\bibfnamefont {N.~B.}\ \bibnamefont {Brookes}}, \bibinfo {author}
  {\bibfnamefont {G.~M.}\ \bibnamefont {{De Luca}}}, \bibinfo {author}
  {\bibfnamefont {A.}~\bibnamefont {Frano}}, \bibinfo {author} {\bibfnamefont
  {D.~G.}\ \bibnamefont {Hawthorn}}, \bibinfo {author} {\bibfnamefont
  {F.}~\bibnamefont {He}}, \bibinfo {author} {\bibfnamefont {T.}~\bibnamefont
  {Loew}}, \bibinfo {author} {\bibfnamefont {M.~M.}\ \bibnamefont {Sala}},
  \bibinfo {author} {\bibfnamefont {D.~C.}\ \bibnamefont {Peets}}, \bibinfo
  {author} {\bibfnamefont {M.}~\bibnamefont {Salluzzo}}, \bibinfo {author}
  {\bibfnamefont {E.}~\bibnamefont {Schierle}}, \bibinfo {author}
  {\bibfnamefont {R.}~\bibnamefont {Sutarto}}, \bibinfo {author} {\bibfnamefont
  {G.~A.}\ \bibnamefont {Sawatzky}}, \bibinfo {author} {\bibfnamefont
  {E.}~\bibnamefont {Weschke}}, \bibinfo {author} {\bibfnamefont
  {B.}~\bibnamefont {Keimer}}, \ and\ \bibinfo {author} {\bibfnamefont
  {L.}~\bibnamefont {Braicovich}},\ }\href {\doibase 10.1126/science.1223532}
  {\bibfield  {journal} {\bibinfo  {journal} {Science}\ }\textbf {\bibinfo
  {volume} {337}},\ \bibinfo {pages} {821} (\bibinfo {year}
  {2012})}\BibitemShut {NoStop}%
\bibitem [{\citenamefont {Chang}\ \emph {et~al.}(2012)\citenamefont {Chang},
  \citenamefont {Blackburn}, \citenamefont {Holmes}, \citenamefont
  {Christensen}, \citenamefont {Larsen}, \citenamefont {Mesot}, \citenamefont
  {Liang}, \citenamefont {Bonn}, \citenamefont {Hardy}, \citenamefont
  {Watenphul}, \citenamefont {Zimmermann}, \citenamefont {Forgan},\ and\
  \citenamefont {Hayden}}]{Chang2012}%
  \BibitemOpen
  \bibfield  {author} {\bibinfo {author} {\bibfnamefont {J.}~\bibnamefont
  {Chang}}, \bibinfo {author} {\bibfnamefont {E.}~\bibnamefont {Blackburn}},
  \bibinfo {author} {\bibfnamefont {A.~T.}\ \bibnamefont {Holmes}}, \bibinfo
  {author} {\bibfnamefont {N.~B.}\ \bibnamefont {Christensen}}, \bibinfo
  {author} {\bibfnamefont {J.}~\bibnamefont {Larsen}}, \bibinfo {author}
  {\bibfnamefont {J.}~\bibnamefont {Mesot}}, \bibinfo {author} {\bibfnamefont
  {R.}~\bibnamefont {Liang}}, \bibinfo {author} {\bibfnamefont {D.~A.}\
  \bibnamefont {Bonn}}, \bibinfo {author} {\bibfnamefont {W.~N.}\ \bibnamefont
  {Hardy}}, \bibinfo {author} {\bibfnamefont {A.}~\bibnamefont {Watenphul}},
  \bibinfo {author} {\bibfnamefont {M.~V.}\ \bibnamefont {Zimmermann}},
  \bibinfo {author} {\bibfnamefont {E.~M.}\ \bibnamefont {Forgan}}, \ and\
  \bibinfo {author} {\bibfnamefont {S.~M.}\ \bibnamefont {Hayden}},\ }\href
  {\doibase 10.1038/nphys2456} {\bibfield  {journal} {\bibinfo  {journal} {Nat.
  Phys.}\ }\textbf {\bibinfo {volume} {8}},\ \bibinfo {pages} {871} (\bibinfo
  {year} {2012})}\BibitemShut {NoStop}%
\bibitem [{\citenamefont {Achkar}\ \emph {et~al.}(2012)\citenamefont {Achkar},
  \citenamefont {Sutarto}, \citenamefont {Mao}, \citenamefont {He},
  \citenamefont {Frano}, \citenamefont {Blanco-Canosa}, \citenamefont {{Le
  Tacon}}, \citenamefont {Ghiringhelli}, \citenamefont {Braicovich},
  \citenamefont {Minola}, \citenamefont {{Moretti Sala}}, \citenamefont
  {Mazzoli}, \citenamefont {Liang}, \citenamefont {Bonn}, \citenamefont
  {Hardy}, \citenamefont {Keimer}, \citenamefont {Sawatzky},\ and\
  \citenamefont {Hawthorn}}]{Achkar2012}%
  \BibitemOpen
  \bibfield  {author} {\bibinfo {author} {\bibfnamefont {A.~J.}\ \bibnamefont
  {Achkar}}, \bibinfo {author} {\bibfnamefont {R.}~\bibnamefont {Sutarto}},
  \bibinfo {author} {\bibfnamefont {X.}~\bibnamefont {Mao}}, \bibinfo {author}
  {\bibfnamefont {F.}~\bibnamefont {He}}, \bibinfo {author} {\bibfnamefont
  {A.}~\bibnamefont {Frano}}, \bibinfo {author} {\bibfnamefont
  {S.}~\bibnamefont {Blanco-Canosa}}, \bibinfo {author} {\bibfnamefont
  {M.}~\bibnamefont {{Le Tacon}}}, \bibinfo {author} {\bibfnamefont
  {G.}~\bibnamefont {Ghiringhelli}}, \bibinfo {author} {\bibfnamefont
  {L.}~\bibnamefont {Braicovich}}, \bibinfo {author} {\bibfnamefont
  {M.}~\bibnamefont {Minola}}, \bibinfo {author} {\bibfnamefont
  {M.}~\bibnamefont {{Moretti Sala}}}, \bibinfo {author} {\bibfnamefont
  {C.}~\bibnamefont {Mazzoli}}, \bibinfo {author} {\bibfnamefont
  {R.}~\bibnamefont {Liang}}, \bibinfo {author} {\bibfnamefont {D.~A.}\
  \bibnamefont {Bonn}}, \bibinfo {author} {\bibfnamefont {W.~N.}\ \bibnamefont
  {Hardy}}, \bibinfo {author} {\bibfnamefont {B.}~\bibnamefont {Keimer}},
  \bibinfo {author} {\bibfnamefont {G.~A.}\ \bibnamefont {Sawatzky}}, \ and\
  \bibinfo {author} {\bibfnamefont {D.~G.}\ \bibnamefont {Hawthorn}},\ }\href
  {\doibase 10.1103/PhysRevLett.109.167001} {\bibfield  {journal} {\bibinfo
  {journal} {Phys. Rev. Lett.}\ }\textbf {\bibinfo {volume} {109}},\ \bibinfo
  {pages} {167001} (\bibinfo {year} {2012})}\BibitemShut {NoStop}%
\bibitem [{\citenamefont {Coslovich}\ \emph {et~al.}(2013)\citenamefont
  {Coslovich}, \citenamefont {Giannetti}, \citenamefont {Cilento},
  \citenamefont {{Dal Conte}}, \citenamefont {Abebaw}, \citenamefont {Bossini},
  \citenamefont {Ferrini}, \citenamefont {Eisaki}, \citenamefont {Greven},
  \citenamefont {Damascelli},\ and\ \citenamefont
  {Parmigiani}}]{Coslovich2013}%
  \BibitemOpen
  \bibfield  {author} {\bibinfo {author} {\bibfnamefont {G.}~\bibnamefont
  {Coslovich}}, \bibinfo {author} {\bibfnamefont {C.}~\bibnamefont
  {Giannetti}}, \bibinfo {author} {\bibfnamefont {F.}~\bibnamefont {Cilento}},
  \bibinfo {author} {\bibfnamefont {S.}~\bibnamefont {{Dal Conte}}}, \bibinfo
  {author} {\bibfnamefont {T.}~\bibnamefont {Abebaw}}, \bibinfo {author}
  {\bibfnamefont {D.}~\bibnamefont {Bossini}}, \bibinfo {author} {\bibfnamefont
  {G.}~\bibnamefont {Ferrini}}, \bibinfo {author} {\bibfnamefont
  {H.}~\bibnamefont {Eisaki}}, \bibinfo {author} {\bibfnamefont
  {M.}~\bibnamefont {Greven}}, \bibinfo {author} {\bibfnamefont
  {A.}~\bibnamefont {Damascelli}}, \ and\ \bibinfo {author} {\bibfnamefont
  {F.}~\bibnamefont {Parmigiani}},\ }\href {\doibase
  10.1103/PhysRevLett.110.107003} {\bibfield  {journal} {\bibinfo  {journal}
  {Phys. Rev. Lett.}\ }\textbf {\bibinfo {volume} {110}},\ \bibinfo {pages}
  {107003} (\bibinfo {year} {2013})}\BibitemShut {NoStop}%
\bibitem [{\citenamefont {Fujita}\ \emph {et~al.}(2014)\citenamefont {Fujita},
  \citenamefont {Hamidian}, \citenamefont {Edkins}, \citenamefont {Kim},
  \citenamefont {Kohsaka}, \citenamefont {Azuma}, \citenamefont {Takano},
  \citenamefont {Takagi}, \citenamefont {Eisaki}, \citenamefont {Uchida},
  \citenamefont {Allais}, \citenamefont {Lawler}, \citenamefont {Kim},
  \citenamefont {Sachdev},\ and\ \citenamefont {Davis}}]{Fujita2014}%
  \BibitemOpen
  \bibfield  {author} {\bibinfo {author} {\bibfnamefont {K.}~\bibnamefont
  {Fujita}}, \bibinfo {author} {\bibfnamefont {M.~H.}\ \bibnamefont
  {Hamidian}}, \bibinfo {author} {\bibfnamefont {S.~D.}\ \bibnamefont
  {Edkins}}, \bibinfo {author} {\bibfnamefont {C.~K.}\ \bibnamefont {Kim}},
  \bibinfo {author} {\bibfnamefont {Y.}~\bibnamefont {Kohsaka}}, \bibinfo
  {author} {\bibfnamefont {M.}~\bibnamefont {Azuma}}, \bibinfo {author}
  {\bibfnamefont {M.}~\bibnamefont {Takano}}, \bibinfo {author} {\bibfnamefont
  {H.}~\bibnamefont {Takagi}}, \bibinfo {author} {\bibfnamefont
  {H.}~\bibnamefont {Eisaki}}, \bibinfo {author} {\bibfnamefont {S.-i.}\
  \bibnamefont {Uchida}}, \bibinfo {author} {\bibfnamefont {A.}~\bibnamefont
  {Allais}}, \bibinfo {author} {\bibfnamefont {M.~J.}\ \bibnamefont {Lawler}},
  \bibinfo {author} {\bibfnamefont {E.-A.}\ \bibnamefont {Kim}}, \bibinfo
  {author} {\bibfnamefont {S.}~\bibnamefont {Sachdev}}, \ and\ \bibinfo
  {author} {\bibfnamefont {J.~C.~S.}\ \bibnamefont {Davis}},\ }\href {\doibase
  10.1073/pnas.1406297111} {\bibfield  {journal} {\bibinfo  {journal} {Proc.
  Natl. Acad. Sci.}\ }\textbf {\bibinfo {volume} {111}},\ \bibinfo {pages}
  {E3026} (\bibinfo {year} {2014})}\BibitemShut {NoStop}%
\bibitem [{\citenamefont {Comin}\ \emph {et~al.}(2015)\citenamefont {Comin},
  \citenamefont {Sutarto}, \citenamefont {He}, \citenamefont {Neto},
  \citenamefont {Chauviere}, \citenamefont {Frano}, \citenamefont {Liang},
  \citenamefont {Hardy}, \citenamefont {Bonn}, \citenamefont {Yoshida},
  \citenamefont {Eisaki}, \citenamefont {Hoffman}, \citenamefont {Keimer},
  \citenamefont {Sawatzky},\ and\ \citenamefont {Damascelli}}]{Comin2014}%
  \BibitemOpen
  \bibfield  {author} {\bibinfo {author} {\bibfnamefont {R.}~\bibnamefont
  {Comin}}, \bibinfo {author} {\bibfnamefont {R.}~\bibnamefont {Sutarto}},
  \bibinfo {author} {\bibfnamefont {F.}~\bibnamefont {He}}, \bibinfo {author}
  {\bibfnamefont {E.~D.~S.}\ \bibnamefont {Neto}}, \bibinfo {author}
  {\bibfnamefont {L.}~\bibnamefont {Chauviere}}, \bibinfo {author}
  {\bibfnamefont {A.}~\bibnamefont {Frano}}, \bibinfo {author} {\bibfnamefont
  {R.}~\bibnamefont {Liang}}, \bibinfo {author} {\bibfnamefont {W.~N.}\
  \bibnamefont {Hardy}}, \bibinfo {author} {\bibfnamefont {D.~A.}\ \bibnamefont
  {Bonn}}, \bibinfo {author} {\bibfnamefont {Y.}~\bibnamefont {Yoshida}},
  \bibinfo {author} {\bibfnamefont {H.}~\bibnamefont {Eisaki}}, \bibinfo
  {author} {\bibfnamefont {J.~E.}\ \bibnamefont {Hoffman}}, \bibinfo {author}
  {\bibfnamefont {B.}~\bibnamefont {Keimer}}, \bibinfo {author} {\bibfnamefont
  {G.~a.}\ \bibnamefont {Sawatzky}}, \ and\ \bibinfo {author} {\bibfnamefont
  {A.}~\bibnamefont {Damascelli}},\ }\href {\doibase 10.1038/nmat4295}
  {\bibfield  {journal} {\bibinfo  {journal} {Nat. Mater.}\ }\textbf {\bibinfo
  {volume} {14}},\ \bibinfo {pages} {796} (\bibinfo {year} {2015})}\BibitemShut
  {NoStop}%
\bibitem [{\citenamefont {Sau}\ and\ \citenamefont {Sachdev}(2014)}]{Sau2013}%
  \BibitemOpen
  \bibfield  {author} {\bibinfo {author} {\bibfnamefont {J.~D.}\ \bibnamefont
  {Sau}}\ and\ \bibinfo {author} {\bibfnamefont {S.}~\bibnamefont {Sachdev}},\
  }\href {\doibase 10.1103/PhysRevB.89.075129} {\bibfield  {journal} {\bibinfo
  {journal} {Phys. Rev. B}\ }\textbf {\bibinfo {volume} {89}},\ \bibinfo
  {pages} {075129} (\bibinfo {year} {2014})}\BibitemShut {NoStop}%
\bibitem [{\citenamefont {Sachdev}\ and\ \citenamefont {{La
  Placa}}(2013)}]{Sachdev2013}%
  \BibitemOpen
  \bibfield  {author} {\bibinfo {author} {\bibfnamefont {S.}~\bibnamefont
  {Sachdev}}\ and\ \bibinfo {author} {\bibfnamefont {R.}~\bibnamefont {{La
  Placa}}},\ }\href {\doibase 10.1103/PhysRevLett.111.027202} {\bibfield
  {journal} {\bibinfo  {journal} {Phys. Rev. Lett.}\ }\textbf {\bibinfo
  {volume} {111}},\ \bibinfo {pages} {027202} (\bibinfo {year}
  {2013})}\BibitemShut {NoStop}%
\bibitem [{\citenamefont {Wang}\ and\ \citenamefont
  {Chubukov}(2014)}]{Wang2014a}%
  \BibitemOpen
  \bibfield  {author} {\bibinfo {author} {\bibfnamefont {Y.}~\bibnamefont
  {Wang}}\ and\ \bibinfo {author} {\bibfnamefont {A.~V.}\ \bibnamefont
  {Chubukov}},\ }\href {\doibase 10.1103/PhysRevB.90.035149} {\bibfield
  {journal} {\bibinfo  {journal} {Phys. Rev. B}\ }\textbf {\bibinfo {volume}
  {90}},\ \bibinfo {pages} {035149} (\bibinfo {year} {2014})}\BibitemShut
  {NoStop}%
\bibitem [{\citenamefont {Patel}\ and\ \citenamefont
  {Eberlein}(2016)}]{Patel2016a}%
  \BibitemOpen
  \bibfield  {author} {\bibinfo {author} {\bibfnamefont {A.~A.}\ \bibnamefont
  {Patel}}\ and\ \bibinfo {author} {\bibfnamefont {A.}~\bibnamefont
  {Eberlein}},\ }\href {\doibase 10.1103/PhysRevB.93.195139} {\bibfield
  {journal} {\bibinfo  {journal} {Phys. Rev. B}\ }\textbf {\bibinfo {volume}
  {93}},\ \bibinfo {pages} {195139} (\bibinfo {year} {2016})}\BibitemShut
  {NoStop}%
\bibitem [{\citenamefont {Raines}\ \emph {et~al.}(2015)\citenamefont {Raines},
  \citenamefont {Stanev},\ and\ \citenamefont {Galitski}}]{Raines2015a}%
  \BibitemOpen
  \bibfield  {author} {\bibinfo {author} {\bibfnamefont {Z.~M.}\ \bibnamefont
  {Raines}}, \bibinfo {author} {\bibfnamefont {V.}~\bibnamefont {Stanev}}, \
  and\ \bibinfo {author} {\bibfnamefont {V.~M.}\ \bibnamefont {Galitski}},\
  }\href {\doibase 10.1103/PhysRevB.91.184506} {\bibfield  {journal} {\bibinfo
  {journal} {Phys. Rev. B}\ }\textbf {\bibinfo {volume} {91}},\ \bibinfo
  {pages} {184506} (\bibinfo {year} {2015})}\BibitemShut {NoStop}%
\bibitem [{\citenamefont {H{\"{o}}ppner}\ \emph {et~al.}(2015)\citenamefont
  {H{\"{o}}ppner}, \citenamefont {Zhu}, \citenamefont {Rexin}, \citenamefont
  {Cavalleri},\ and\ \citenamefont {Mathey}}]{Hoppner2014a}%
  \BibitemOpen
  \bibfield  {author} {\bibinfo {author} {\bibfnamefont {R.}~\bibnamefont
  {H{\"{o}}ppner}}, \bibinfo {author} {\bibfnamefont {B.}~\bibnamefont {Zhu}},
  \bibinfo {author} {\bibfnamefont {T.}~\bibnamefont {Rexin}}, \bibinfo
  {author} {\bibfnamefont {A.}~\bibnamefont {Cavalleri}}, \ and\ \bibinfo
  {author} {\bibfnamefont {L.}~\bibnamefont {Mathey}},\ }\href {\doibase
  10.1103/PhysRevB.91.104507} {\bibfield  {journal} {\bibinfo  {journal} {Phys.
  Rev. B}\ }\textbf {\bibinfo {volume} {91}},\ \bibinfo {pages} {11} (\bibinfo
  {year} {2015})}\BibitemShut {NoStop}%
\bibitem [{\citenamefont {Kivelson}\ \emph {et~al.}(1990)\citenamefont
  {Kivelson}, \citenamefont {Emery},\ and\ \citenamefont {Lin}}]{Kivelson1990}%
  \BibitemOpen
  \bibfield  {author} {\bibinfo {author} {\bibfnamefont {S.~A.}\ \bibnamefont
  {Kivelson}}, \bibinfo {author} {\bibfnamefont {V.~J.}\ \bibnamefont {Emery}},
  \ and\ \bibinfo {author} {\bibfnamefont {H.~Q.}\ \bibnamefont {Lin}},\ }\href
  {\doibase 10.1103/PhysRevB.42.6523} {\bibfield  {journal} {\bibinfo
  {journal} {Phys. Rev. B}\ }\textbf {\bibinfo {volume} {42}},\ \bibinfo
  {pages} {6523} (\bibinfo {year} {1990})}\BibitemShut {NoStop}%
\bibitem [{\citenamefont {Dagotto}\ and\ \citenamefont
  {Riera}(1992)}]{Dagotto1992}%
  \BibitemOpen
  \bibfield  {author} {\bibinfo {author} {\bibfnamefont {E.}~\bibnamefont
  {Dagotto}}\ and\ \bibinfo {author} {\bibfnamefont {J.}~\bibnamefont
  {Riera}},\ }\href {\doibase 10.1103/PhysRevB.46.12084} {\bibfield  {journal}
  {\bibinfo  {journal} {Phys. Rev. B}\ }\textbf {\bibinfo {volume} {46}},\
  \bibinfo {pages} {12084} (\bibinfo {year} {1992})}\BibitemShut {NoStop}%
\bibitem [{\citenamefont {Allais}\ \emph
  {et~al.}(2014{\natexlab{a}})\citenamefont {Allais}, \citenamefont {Bauer},\
  and\ \citenamefont {Sachdev}}]{Allais2014}%
  \BibitemOpen
  \bibfield  {author} {\bibinfo {author} {\bibfnamefont {A.}~\bibnamefont
  {Allais}}, \bibinfo {author} {\bibfnamefont {J.}~\bibnamefont {Bauer}}, \
  and\ \bibinfo {author} {\bibfnamefont {S.}~\bibnamefont {Sachdev}},\ }\href
  {\doibase 10.1103/PhysRevB.90.155114} {\bibfield  {journal} {\bibinfo
  {journal} {Phys. Rev. B}\ }\textbf {\bibinfo {volume} {90}},\ \bibinfo
  {pages} {155114} (\bibinfo {year} {2014}{\natexlab{a}})}\BibitemShut
  {NoStop}%
\bibitem [{Note1()}]{Note1}%
  \BibitemOpen
  \bibinfo {note} {In this work we used $t_1=\SI {430}{meV}$, $t_2 = -0.32t_1$,
  $t_3 = -0.5t_2$, and $\mu =-1.1856t_1$.}\BibitemShut {Stop}%
\bibitem [{\citenamefont {Chakravarty}\ \emph {et~al.}(1993)\citenamefont
  {Chakravarty}, \citenamefont {Sudbo}, \citenamefont {Anderson},\ and\
  \citenamefont {Strong}}]{chakravarty_interlayer_1993}%
  \BibitemOpen
  \bibfield  {author} {\bibinfo {author} {\bibfnamefont {S.}~\bibnamefont
  {Chakravarty}}, \bibinfo {author} {\bibfnamefont {A.}~\bibnamefont {Sudbo}},
  \bibinfo {author} {\bibfnamefont {P.~W.}\ \bibnamefont {Anderson}}, \ and\
  \bibinfo {author} {\bibfnamefont {S.}~\bibnamefont {Strong}},\ }\href
  {\doibase 10.1126/science.261.5119.337} {\bibfield  {journal} {\bibinfo
  {journal} {Science}\ }\textbf {\bibinfo {volume} {261}},\ \bibinfo {pages}
  {337} (\bibinfo {year} {1993})}\BibitemShut {NoStop}%
\bibitem [{\citenamefont {Xiang}\ and\ \citenamefont
  {Hardy}(2000)}]{Xiang2000}%
  \BibitemOpen
  \bibfield  {author} {\bibinfo {author} {\bibfnamefont {T.}~\bibnamefont
  {Xiang}}\ and\ \bibinfo {author} {\bibfnamefont {W.~N.}\ \bibnamefont
  {Hardy}},\ }\href {\doibase 10.1103/PhysRevB.63.024506} {\bibfield  {journal}
  {\bibinfo  {journal} {Phys. Rev. B}\ }\textbf {\bibinfo {volume} {63}},\
  \bibinfo {pages} {024506} (\bibinfo {year} {2000})}\BibitemShut {NoStop}%
\bibitem [{\citenamefont {Thomson}\ and\ \citenamefont
  {Sachdev}(2015)}]{Thomson2014}%
  \BibitemOpen
  \bibfield  {author} {\bibinfo {author} {\bibfnamefont {A.}~\bibnamefont
  {Thomson}}\ and\ \bibinfo {author} {\bibfnamefont {S.}~\bibnamefont
  {Sachdev}},\ }\href {\doibase 10.1103/PhysRevB.91.115142} {\bibfield
  {journal} {\bibinfo  {journal} {Phys. Rev. B}\ }\textbf {\bibinfo {volume}
  {91}},\ \bibinfo {pages} {115142} (\bibinfo {year} {2015})}\BibitemShut
  {NoStop}%
\bibitem [{\citenamefont {Chowdhury}\ and\ \citenamefont
  {Sachdev}(2014)}]{Chowdhury2014}%
  \BibitemOpen
  \bibfield  {author} {\bibinfo {author} {\bibfnamefont {D.}~\bibnamefont
  {Chowdhury}}\ and\ \bibinfo {author} {\bibfnamefont {S.}~\bibnamefont
  {Sachdev}},\ }\href {\doibase 10.1103/PhysRevB.90.134516} {\bibfield
  {journal} {\bibinfo  {journal} {Phys. Rev. B}\ }\textbf {\bibinfo {volume}
  {90}},\ \bibinfo {pages} {134516} (\bibinfo {year} {2014})}\BibitemShut
  {NoStop}%
\bibitem [{\citenamefont {Allais}\ \emph
  {et~al.}(2014{\natexlab{b}})\citenamefont {Allais}, \citenamefont {Bauer},\
  and\ \citenamefont {Sachdev}}]{Allais2014a}%
  \BibitemOpen
  \bibfield  {author} {\bibinfo {author} {\bibfnamefont {A.}~\bibnamefont
  {Allais}}, \bibinfo {author} {\bibfnamefont {J.}~\bibnamefont {Bauer}}, \
  and\ \bibinfo {author} {\bibfnamefont {S.}~\bibnamefont {Sachdev}},\ }\href
  {\doibase 10.1007/s12648-014-0488-4} {\bibfield  {journal} {\bibinfo
  {journal} {Indian J. Phys.}\ }\textbf {\bibinfo {volume} {88}},\ \bibinfo
  {pages} {905} (\bibinfo {year} {2014}{\natexlab{b}})}\BibitemShut {NoStop}%
\bibitem [{\citenamefont {Nyhus}\ \emph {et~al.}(1994)\citenamefont {Nyhus},
  \citenamefont {Karlow}, \citenamefont {Cooper}, \citenamefont {Veal},\ and\
  \citenamefont {Paulikas}}]{Nyhus1994}%
  \BibitemOpen
  \bibfield  {author} {\bibinfo {author} {\bibfnamefont {P.}~\bibnamefont
  {Nyhus}}, \bibinfo {author} {\bibfnamefont {M.~A.}\ \bibnamefont {Karlow}},
  \bibinfo {author} {\bibfnamefont {S.~L.}\ \bibnamefont {Cooper}}, \bibinfo
  {author} {\bibfnamefont {B.~W.}\ \bibnamefont {Veal}}, \ and\ \bibinfo
  {author} {\bibfnamefont {A.~P.}\ \bibnamefont {Paulikas}},\ }\href {\doibase
  10.1103/PhysRevB.50.13898} {\bibfield  {journal} {\bibinfo  {journal} {Phys.
  Rev. B}\ }\textbf {\bibinfo {volume} {50}},\ \bibinfo {pages} {13898}
  (\bibinfo {year} {1994})}\BibitemShut {NoStop}%
\bibitem [{\citenamefont {Honma}\ and\ \citenamefont {Hor}(2010)}]{Honma2010}%
  \BibitemOpen
  \bibfield  {author} {\bibinfo {author} {\bibfnamefont {T.}~\bibnamefont
  {Honma}}\ and\ \bibinfo {author} {\bibfnamefont {P.~H.}\ \bibnamefont
  {Hor}},\ }\href {\doibase 10.1016/j.ssc.2010.10.003} {\bibfield  {journal}
  {\bibinfo  {journal} {Solid State Commun.}\ }\textbf {\bibinfo {volume}
  {150}},\ \bibinfo {pages} {2314} (\bibinfo {year} {2010})}\BibitemShut
  {NoStop}%
\bibitem [{Note2()}]{Note2}%
  \BibitemOpen
  \bibinfo {note} {The wrapped normal distribution is a straightforward
  extension of the normal distribution to a periodic variable. It is a close
  cousin of the Von Mises distribution, which is the eigendistribution of
  diffusion for a periodic variable with a harmonic confinement but is somewhat
  more analytically convenient. We have explicitly checked that there is no
  qualitative difference between the results for the two
  distributions.}\BibitemShut {Stop}%
\bibitem [{\citenamefont {Denny}\ \emph {et~al.}(2015)\citenamefont {Denny},
  \citenamefont {Clark}, \citenamefont {Laplace}, \citenamefont {Cavalleri},\
  and\ \citenamefont {Jaksch}}]{denny_proposed_2015}%
  \BibitemOpen
  \bibfield  {author} {\bibinfo {author} {\bibfnamefont {S.}~\bibnamefont
  {Denny}}, \bibinfo {author} {\bibfnamefont {S.}~\bibnamefont {Clark}},
  \bibinfo {author} {\bibfnamefont {Y.}~\bibnamefont {Laplace}}, \bibinfo
  {author} {\bibfnamefont {A.}~\bibnamefont {Cavalleri}}, \ and\ \bibinfo
  {author} {\bibfnamefont {D.}~\bibnamefont {Jaksch}},\ }\href {\doibase
  10.1103/PhysRevLett.114.137001} {\bibfield  {journal} {\bibinfo  {journal}
  {Phys. Rev. Lett.}\ }\textbf {\bibinfo {volume} {114}},\ \bibinfo {pages}
  {137001} (\bibinfo {year} {2015})}\BibitemShut {NoStop}%
\end{thebibliography}%
\end{document}